\documentclass[bibyear]{aa}  

\usepackage{graphicx}
\usepackage{txfonts}
\begin{document}

\title{Cloud-cloud collision as origin of the G31.41+0.31 massive protocluster}


   \author{M.\ T.\ Beltr\'an\inst{1}, 
V.\ M.\ Rivilla\inst{2, 1}, M.\ S.\ N.\ Kumar\inst{3}, R.\ Cesaroni\inst{1}, D.\ Galli\inst{1}
}

   \institute{INAF-Osservatorio Astrofisico di Arcetri, Largo E.\ Fermi 5,
I-50125 Firenze, Italy\\
              \email{maria.beltranl@inaf.it}
         \and
Centro de Astrobiolog\'{i}a (CSIC-INTA), Ctra.\ de Ajalvir Km. 4, Torrej\'on de Ardoz, 28850, Madrid, Spain \\
\and
Instituto de Astrof\'{i}sica e Ci\^encias do Espa\c{c}o, Universidade do Porto, CAUP, Rua das Estrelas, 4150-762, Porto, Portugal \\
}
 
   \date{Received ; accepted }
   
   \titlerunning{Large-scale kinematics in G31.41+0.31}
\authorrunning{Beltr\'an et al.}

 \abstract{The G31.41+0.31 (G31) hot molecular core (HMC) is a high-mass protocluster showing accelerated infall and rotational spin-up that is well studied at high-angular resolution. To complement the accurate view of the small scale in G31, we have traced the kinematics of the large-scale material by carrying out N$_2$H$^+$\,(1--0) observations with the IRAM 30m telescope of an area of $\sim$6$\times6$\,arcmin$^2$ around the HMC. The N$_2$H$^+$ observations have revealed a large-scale (5\,pc) hub-filament system (HFS) composed by at least four filamentary arms and a NNE--SSW velocity gradient ($\sim$0.4\,km/s/pc) between the northern and southern filaments. The linewidth increases towards the hub at the center of the HFS reaching values of 2.5--3\,km\,s$^{-1}$ in the central 1\,pc. The origin of the large-scale velocity gradient is likely cloud-cloud collision. In this scenario, the filaments in G31 would have formed by compression resulting from the collision and the rotation of the HMC observed at scales of 1000\,au would have been induced by shear caused by the cloud-cloud collision at scales of a few pc. We conclude that G31 represents a HFS in a compressed layer with an orthogonal orientation to the plane of the sky, and represents a benchmark for the filaments-to-clusters  paradigm of star formation.
}
 

   \maketitle
%

\section{Introduction}
\label{sect-intro}

G31.41+0.31 (G31 hereafter) is a high-mass star-forming region located at 3.75 kpc (Immer et al.~\cite{immer19}) and with a luminosity of $\sim$5$\times10^4\,L_\odot$ that harbors a very chemically rich hot molecular core (HMC) (e.g., Beltr\'an et al.~\cite{beltran09}; Rivilla et al.~\cite{rivilla17}; Mininni et al.~\cite{mininni20}; Colzi et al.~\cite{colzi21}) and an ultra-compact (UC) H{\sc ii} region, located at $\sim$5$''$ from the HMC. The region has been extensively observed at high-angular resolution ($<1''$) with interferometers, and the observations have revealed that: $i)$ the HMC has already fragmented and formed a small protocluster composed of at least four massive sources within 1$''$ (Beltr\'an et al.~\cite{beltran21}). Besides the sources embedded in the Main core of G31, there are six additional millimeter sources in the region, very close to the HMC and located in what appears as streams/filaments of matter pointing to the HMC (Beltr\'an et al.~\cite{beltran21}); $ii)$ the core displays a clear northeast-southwest (NE--SW) velocity gradient, observed in several high-density tracers, that has been interpreted as due to rotation (Beltr\'an et al.~\cite{beltran04}, \cite{beltran05}, \cite{beltran18}; Girart et al.~\cite{girart09}; Cesaroni et al.~\cite{cesa11}); $iii)$ the core is still actively accreting, and the kinematics studied at high-angular resolution ($\sim$0$\farcs2$ or 750\,au) indicates accelerating infall and rotational spin-up (Beltr\'an et al. 2018); and $iv)$ dust polarization observations reveal an hour-glass shaped magnetic field with the symmetry axis oriented perpendicular to the velocity gradient (Girart et al.~\cite{girart09}; Beltr\'an et al.~\cite{beltran19}).

\begin{figure*}
\centerline{\includegraphics[width=11.5cm,angle=90]{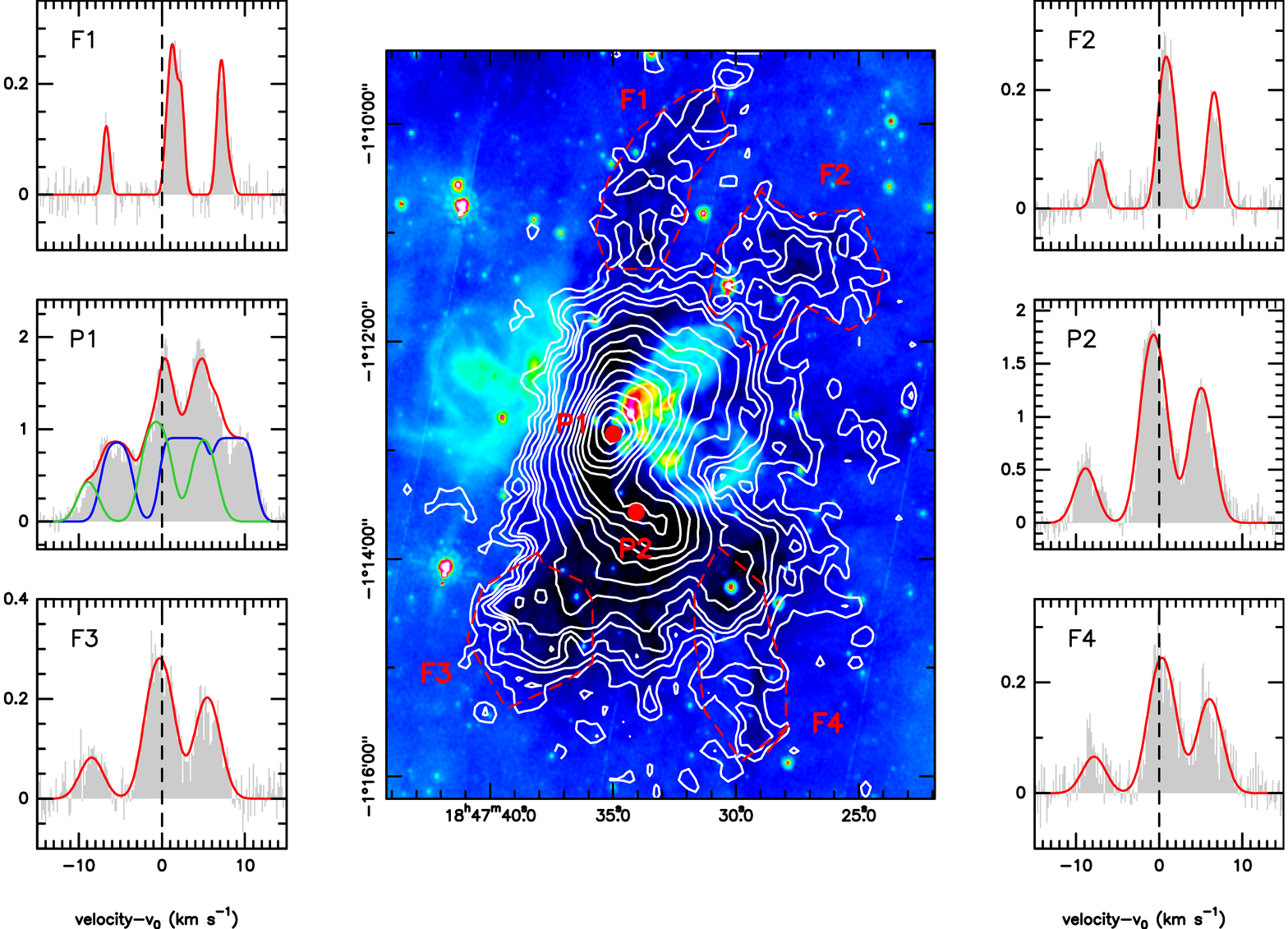}}
\caption{({\it Middle panel}) Total N$_2$H$^+$\,(1--0) emission integrated between 80 and 110\,\ overlaid on the wide-field infrared {\it Spitzer} 8\,$\mu$m image of the G31 star-forming region. The coordinates are J2000 right ascension and declination. The white contours show emission levels from 0.5 to 3 K km~s$^{-1}$ (in steps of 0.5 K km~s$^{-1}$), and from 4 to 24 K km~s$^{-1}$ (in steps of 2 K km~s$^{-1}$). The red dots and the red dashed areas denote selected positions and regions, respectively, whose spectra are shown in the left and right panels. ({\it Left and right panels}): N$_2$H$^+$\,(1--0) spectra (gray histograms) toward selected positions and regions. The red curves show the best LTE fit obtained with MADCUBA. Blue and green curves in P1 show the two velocity components that better fit the data. 
The vertical dashed line corresponds to $v$-$v_{\rm 0}$=0, where $v_{\rm 0}$ is the systemic velocity of 96.5 km~s$^{-1}$ (Beltr\'an et al.~\cite{beltran18}). The y-axis scale shows the main beam temperature, $T_{\rm B}$, in K, at the peak positions (P1 and P2) or averaged within the regions (F1, F2, F2, and F4). }
\label{fig-n2h+}
\end{figure*} 

The wide-field infrared view of the G31 region as observed with {\it Spitzer} at 8\,$\mu$m evidences that this well-studied (at small scales) HMC is located at the junction of multiple filaments (Fig.~\ref{fig-n2h+}). This together with the existence of a protocluster of millimeter sources in G31 (Beltr\'an et al.~\cite{beltran21}) suggests that this massive core might represent the young protostellar hub of a large hub-filament system (HFS), similar to those described by the Filaments to Clusters (F2C) scheme of Kumar et al.~(\cite{kumar20}). This paradigm proposes 4 stages for massive star formation (see Fig.~14 of Kumar et al.~\cite{kumar20}): I)
dense filaments, formed via mechanisms such as
cloud–cloud collision, move toward each other and set up
the initial conditions for the formation of a HFS. II) Filaments collide and form a hub. The hub gains a twist as the overlap point is different from the center of mass and this gives rise to an initial angular momentum. The resulting spin can eventually flatten the hub. III)  Column density amplification in the hub triggers star formation and produces a gravitational potential difference between the hub and the filament, which can drive longitudinal flows within the filaments directed toward the hub. IV) Radiation pressure and ionisation feedback from the newly formed OB stars  escapes through the inter-filamentary cavities by punching holes in the flattened hub. The filaments are eroded by the expanding radiation bubbles creating pillars, while a mass-segregated embedded cluster is formed in the hub. 

Although G31 is one of the most studied HMCs at high-angular resolution, little is known about the material and kinematics of the large-scale region besides the wide-field infrared images that clearly show the presence of dark lanes and filaments that point to G31. Up to now there were no large-scale emission line observations to identify signatures of matter accreting onto the central core from the larger-scale cloud, which 
according to the F2C paradigm of Kumar et al.~(\cite{kumar20}), should exist. In fact, such flows have been seen in other massive HFSs such as SDC13 (Peretto et al.~\cite{peretto14}) and G14.2 (Chen et al.~\cite{chen19}). Therefore, to get an accurate view of high-mass star-formation and complete the whole picture, from large (clump) to small (core and disk) scales, we observed the G31 region with the Institute de Radioastronomie Millim\'etrique (IRAM) 30m telescope in N$_2$H$^+$, a typical high-density tracer and commonly used to trace the kinematics of filaments in high- mass star-forming regions (Peretto et al.~\cite{peretto14}; Hacar et al.~\cite{hacar18}). The goal of the observations was to complement the high-resolution study of the small scale in G31 by tracing the kinematics of the large-scale material and by studying its importance on the growth of massive stars in

\section{Observations}
\label{sect-IRAM}

The observations were carried out with the IRAM 30m telescope during two 8\,hr observing runs between July 8 and 10, 2021 (project number 049$-$21).  We used the Eight Mixer Receiver (EMIR) and the Fast Fourier Transform Spectrometer (FTS), centering the lower inner band at the frequency of N$_2$H$^{+}$\,(1--0)  (93.1734035\,GHz). The spectral resolution was 50\,kHz, which translates to
0.161 km s$^{-1}$. The half power beam width (HPBW) was 26$\farcs$4. We used on-the-fly position-switching observing mode to cover a total central area of $\sim$6$\times6$\,arcmin$^2$ plus a few more arcmin$^2$ following the infrared-dark (IR-dark) filaments. The phase reference center of the observations was set to the position $\alpha$(J2000)=18$^{\rm h}$ 47$^{\rm m}$ 34$\fs$315, $\delta$(J2000) $-$01$^{\circ}$12$^{\prime}$45.9$^{\prime\prime}$. The mosaic consisted of 13 different sub-maps with sizes of 120$\times$120 arcsec$^2$. Each sub-map was observed several times and in orthogonal directions. The off position, ($-$885 arcsec, 290.00 arcsec) with respect to the phase center, was chosen because it is not associated with $^{13}$CO emission as seen in the maps of the Boston University-FCRAO Galactic Ring Survey (GRS, Jackson et al.~\cite{jackson06}). The pointing was checked every 1.5\,hr, and focus was checked at the beginning and at the middle of each observing run toward planets and/or bright quasars. The line intensity of the spectra was converted to the main beam temperature $T_{\rm mb}$, which was calculated as: $T_{\rm mb}$=$T_{\rm A}^*\times F_{\rm eff}/B_{\rm eff}$, where $T_{\rm A}^*$ is the antenna temperature, and $F_{\rm eff}$ and $B_{\rm eff}$ are the forward and beam efficiencies\footnote{http://www.iram.
es/IRAMES/mainWiki/Iram30mEfficiencies}, respectively ($F_{\rm eff}$=95 and $B_{\rm eff}$=80) The data were reduced using the GILDAS/CLASS software\footnote{http://www.iram.fr/IRAMFR/GILDAS}. The data of the two observing runs and submaps were combined and Nyquist resampled with a final resolution of 27$\farcs$8.

\begin{figure*}
\centerline{\includegraphics[angle=0,width=11cm,angle=0]{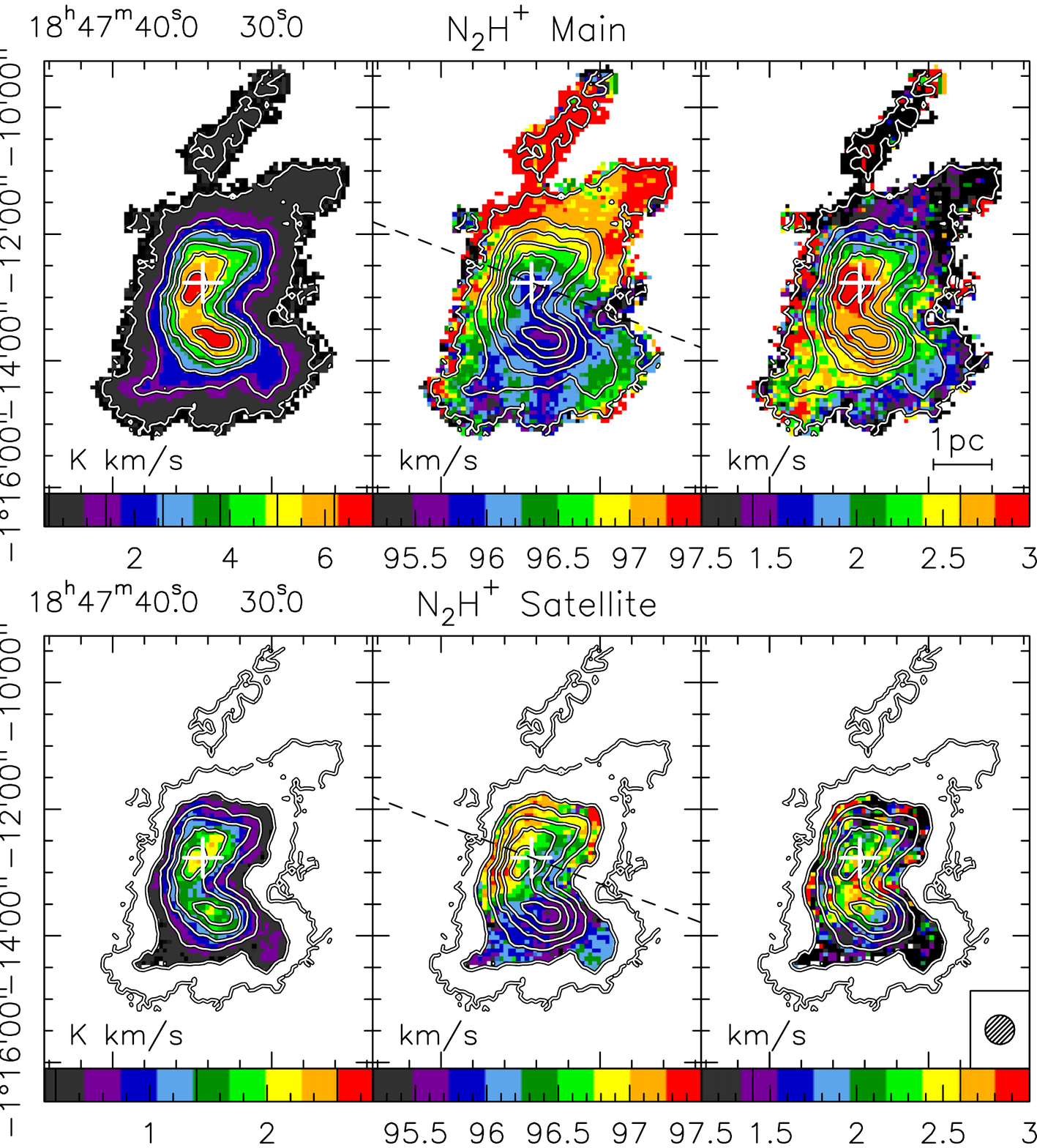}}
\caption{Integrated intensity (moment 0) map of the N$_2$H$^+$\,(1--0) main line centered at 93.1737699\,GHz ({\it white contours}) overlaid on the integrated intensity, line velocity (moment 1), and linewidth (moment 2) maps ({\it colours}) for the main line  ({\it top panels}) and the satellite line centered at 93.1762595\,GHz ({\it bottom panels}).  The line emission has been integrated over the velocity range 93 to 98\,km~s$^{-1}$. The white contours range from 0.2\,K\,km~s$^{-1}$ to 6.2\,K\,km~s$^{-1}$ by 1.2\,K\,km~s$^{-1}$. The white cross indicates the position of the dust continuum emission peak of the G31 HMC. The black dash line in the velocity maps indicates the direction of the NE--SW velocity gradient observed in the  G31 HMC in several high-density tracers at high-angular resolution (e.g., Beltr\'an et al.~\cite{beltran18}). The IRAM\,30m beam is shown in the lower right-hand corner of the bottom right panel.}
\label{fig-moments}
\end{figure*} 

\section{Results and Analysis}

\subsection{N$_2$H$^+$ emission}
\label{sect-n2h+}

Figure~\ref{fig-n2h+} shows the total N$_2$H$^+$\,(1--0) integrated emission overlaid on the wide-field infrared {\it Spitzer} 8\,$\mu$m image of the G31 star-forming region, which shows the IR-dark filamentary cloud surrounding the G31 HMC and giving it the characteristic aspect of a HFS. As seen in this figure, the gas emission coincides perfectly with the IR-dark cloud, in particular with the filamentary structure. Dust continuum emission observed with {\it Herschel} as part of the Hi-GAL key project (Molinari et al.~\cite{molinari10}) from 160\,$\mu$m to 500\,$\mu$m traces the same material as N$_2$H$^+$.

To estimate the physical properties of the gas in the IR-dark cloud, we took N$_2$H$^+$ spectra toward selected regions, which includes regions towards the filaments (F1, F2, F3 and F4), and the positions of the N$_2$H$^+$ emission peaks (P1 and P2, primary and secondary peaks, respectively). The regions studied are shown in Fig.~\ref{fig-n2h+}. We used the SLIM (Spectral Line Identification and Modeling) tool within the MADCUBA package{\footnote{Madrid Data Cube Analysis on ImageJ is a software developed at the Center of Astrobiology (CAB) in Madrid; http://cab.inta-csic.es/madcuba/Portada.html.}} (version 26/07/2021; Mart\'{i}n et al.~\cite{martin19}) to fit the emission toward each region. We performed a Local Thermodynamic Equilibrium (LTE) analysis using SLIM.  As demonstrated by Mart\'{i}n et al.~(\cite{martin19}), the LTE fit of the hyperfine structure of the $J$=1$-$0 transition allows to derive also the excitation temperature, $T_{\rm ex}$, thanks to the different line opacities of the different hyperfine components.
To perform the LTE analysis, we left as free parameters the column density, $N$, $T_{\rm ex}$, the velocity, $v_{\rm LSR}$, and the full width at half maximum (FWHM). 
For the emission of the central region (P1), two different velocity components are needed to fit the emission.
The results of the LTE fits are shown in Fig.~\ref{fig-n2h+} and summarized in Table \ref{table-n2h+} of the Appendix. For the regions of the filaments (F1, F2, F3 and F4), we obtained column densities in the range (3$-$9)$\times$10$^{13}$ cm$^{-2}$, FWHMs of 0.9$-$3.2 km~s$^{-1}$, and low $T_{\rm ex}$ of 3$-$3.5 K. The line opacities ($\tau$) of the main hyperfine transition (1$_{2,3}-$0$_{1,2}$) are 0.3$-$0.6.
For the peak positions of the N$_2$H$^+$ emission (P1 and P2), we derived $T_{\rm ex}$ and FWHM similar to those of the filaments, but significantly higher column densities, (20$-$150)$\times$10$^{13}$ cm$^{-2}$. The line opacity of one of the velocity component of the central region (P1) is $\tau>$7, indicating that this component is heavily optically thick (see blue curve in Fig.~\ref{fig-n2h+}, left panel).

\subsection{N$_2$H$^+$ moment maps}

Figure~\ref{fig-moments} shows the integrated intensity (moment 0), line velocity (moment 1), and linewidth (moment 2) maps of the N$_2$H$^+$\,(1--0) main line centered at 93.1737699\,GHz and the satellite line centered at 93.1762595\,GHz. The line emission has been integrated over the velocity range 93 to 98\,km~s$^{-1}$. As seen in this figure, the integrated emission shows two well-defined emission peaks, one to the north associated with the G31 HMC and the UC\,H{\sc ii} region, and one to the south not associated with known young stellar objects. The satellite line moment 0 map, which is less affected by opacity effects than the main line, shows that the peak associated with on-going star formation, which is the strongest, breaks into two: a peak clearly associated with the HMC, and another slightly to the northwest.

The line velocity maps (Fig.~\ref{fig-moments}) show a clear NNE--SSW velocity gradient across the cloud extending over nearly 5\,pc, from the northern to the southern filaments that suggests global motions in the cloud. The value of the velocity gradient is $\sim$0.4\,km/s/pc. The linewidth toward the position of the G31 HMC/protocluster (P1), i.e., in the inner 1\,pc of the hub region, is  2.5--3\,km~s$^{-1}$ as estimated from the satellite line ($F_1$=0$-$1).  Note that the linewidth toward the HMC estimated from the main line (1$_{2,3}-$0$_{1,2}$) ({\it right upper panel} in Fig.~\ref{fig-moments}) should be taken as an upper limit because this line is optically thick as compared to the satellites. In addition, the main transition is blended with other $F_1$=2$-$1 hyperfine transitions, as shown in Table~\ref{table-n2h+}. The linewidth along the filaments and within the southern emission peak (lacking YSOs; P2) is very small, with values of $\lesssim$1\,km~s$^{-1}$.

\section{Discussion}

\subsection{Kinematics at large scale}

The velocity line maps of N$_2$H$^+$ have revealed the existence of a clear NNE--SSW velocity gradient at large scale in the region (Fig.~\ref{fig-moments}).  The direction of this velocity gradient, especially as seen in the optically thinner satellite line, matches that of the well-known gradient (at scales of $<$0.1\,pc) observed in several high-density tracers, such as CH$_3$CN, at high-angular resolution (better than 1$''$ or 3750\,au) toward the G31 HMC (Beltr\'an et al.~\cite{beltran04}, \cite{beltran05}, \cite{beltran18}; Girart et al.~\cite{girart09}; Cesaroni et al.~\cite{cesa11}) and interpreted as rotation of the core. The direction of the small-scale velocity gradient is indicated with a dashed line in Fig.~\ref{fig-moments}.  We have identified two possible origins for the gradient: rotation and cloud-cloud collision. A gradient of $\sim$0.4\,km/s/pc at scales of $\sim$5\,pc, if interpreted as rotation, would correspond to an angular rotation $\Omega$ of $\sim$1.3$\times10^{-14}$\,s$^{-1}$. This corresponds to a rotation period of $\sim$15$\times10^6$\,yr and would need a central mass of $\sim$600\,$M_\odot$, for a rotation radius of $\sim$2.5\,pc.  Cesaroni et al.~(\cite{cesa11}) have estimated a mass of $\sim$400\,$M_\odot$ for the G31 HMC, which is, however, not concentrated at the center of the cloud. Therefore, although rotation may be supported, it is unlikely that gas on parsec scales has been revolving for about 15 million years around a 600\,$M_\odot$ star cluster concentrated at the center of the system. 

The most likely explanation for the large-scale velocity gradient is cloud-cloud collision. In this scenario, the filaments in G31 visible in absorption in the near-IR and in emission in N$_2$H$^+$ and far-IR continuum would have formed by compression resulting from the collision (e.g., Inoue \& Fukui~\cite{inoue13}), and the rotation of the HMC observed at scales of 1000\,au would have been induced by shear caused by the cloud-cloud collision (Balfour et al.~\cite{balfour15}, \cite{balfour17}). According to Kumar et al.~(\cite{kumar20}), cloud-cloud collision could be the most likely mechanism through which hubs with large networks of filaments form in massive clouds (see also, Balfour et al.~\cite{balfour15}; Fukui et al.~\cite{fukui21}). The material is conveyed from the large-scale could to the hubs through these filaments and this leads to cluster formation, as observed in other high-mass star-forming regions, e.g., SDC13 (Peretto et al.~\cite{peretto14}) and G14.2 (Chen et al.~\cite{chen19}).  In this scenario, the hub gains a
twist as the point where the filaments intersect is different from the center of mass.

The cloud-cloud collision scenario is supported by the fact that the N$_2$H$^+$ spectra show two velocity components toward the HMC (see Fig.~\ref{fig-n2h+} and Table~\ref{table-n2h+}), while only one velocity component is visible at all the other positions. The spatial resolution of the observations here is insufficient to trace the longitudinal flows along the filaments as observed in other high-mass star-forming regions, e.g.,  SDC13 (Peretto et al.~\cite{peretto14}) and G14.2 (Chen et al.~\cite{chen19}). However, one can estimate the mass available in the filaments that could still be incorporated in the central hubs. For this we have used the  N$_2$H$^+$ mean column density of 4.1$\times10^{12}$\,cm$^{-2}$, estimated in an area of $\sim$4600\,arcsec$^2$ for the most clear filament in the region, the northern filament or region A in Fig.~\ref{fig-n2h+} and Table~\ref{table-n2h+}. Assuming an N$_2$H$^+$ abundance of $\sim$2.5$\times10^{-10}$ (Hacar et al.~\cite{hacar17}), the mass available in this filament is $\sim$460\,$M_\odot$. 

\begin{figure}
\centerline{\includegraphics[angle=0,width=10cm,angle=0]{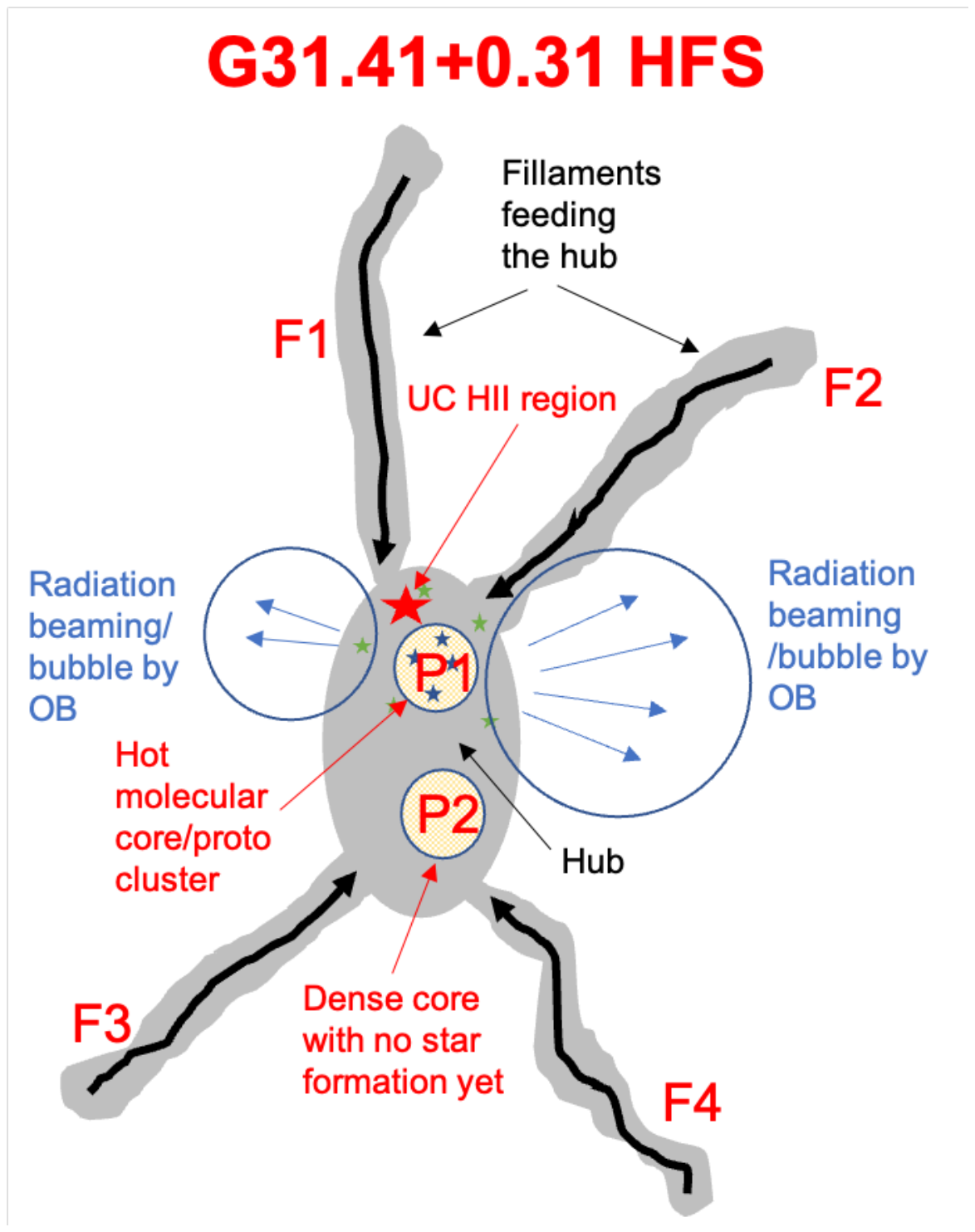}}
\caption{Sketch of the evolutionary stage of G31 within the F2C paradigm (Kumar et al.~\cite{kumar20}). A hub is formed by the junction of filaments. The hub gravitational potential triggers and drives longitudinal flows bringing additional matter and further enhancing the density. Hub fragmentation results in two star formation sites with different ages. Radiation pressure and ionization feedback escapes through the inter-filamentary cavities by punching holes in the flattened hub.} 
\label{fig-sketch}
\end{figure}

\subsection{G31: A benchmark for the F2C paradigm}

N$_2$H$^+$ and far-IR observations suggest that G31 could represent a typical example of HFS, where star-formation is explained in terms of the F2C paradigm postulated by Kumar et al.~(\cite{kumar20}, \cite{kumar21}). In this scenario, the elongated and flattened hub is located at the center of two nearly equal sized lobes of the bipolar shaped radiation bubble visible in the infrared (Fig.~\ref{fig-n2h+}), and is fed by material conveyed though the filaments (see the sketch in Fig.~\ref{fig-sketch}). The HFS shows two  N$_2$H$^+$ emission peaks located on either side of the hub center (Fig.~\ref{fig-moments}), one located at the position of the HMC G31, where a massive protocluster has already formed (Beltrán et al. 2021), and the other to the south, which could be the next site of massive star formation. These dense cores have also been observed in dust continuum emission with {\it Herschel} from 160\,$\mu$m to 500\,$\mu$m, and are similar to the two centers of activity in a flattened hub described by the F2C paradigm (Kumar et al.~\cite{kumar21}), where one center is at an earlier evolutionary stage than the other. 

The fact that the HMC with its associated massive protocluster is located close ($\sim$5$''$) to a more evolved UC\,H{\sc ii} region, ionized by an O6 star, suggests that G31 could be in transition between Stage\,III and IV  (Fig.~\ref{fig-sketch}) of the evolutionary scenario proposed by Kumar et al.~(\cite{kumar20}). This hypothesis is supported by the fact that infrared emission bubbles (which could indicate UV radiation escaping through the inter-filamentary cavities) are already visible  (Fig.~\ref{fig-n2h+}) as in Stage\,IV. However, the fact there are still no pillars visible suggests that this hub is still in a phase prior to Stage\,IV. Taking into account the morphology of the region, G31 could represent an edge-on view of the HFSs described by Kumar et al.~(\cite{kumar20}) in an intermediate phase stage between III and IV. In contrast, Mon\,R2 (Kumar et al.~\cite{kumar21}) represents a face-on view of a system similar to G31.

Dust polarization observations of G31 carried out with SMA at 1$''$ (Girart et al.~\cite{girart09}) and ALMA at $0.02''$ (Beltr\'an et al.~\cite{beltran19}) have revealed an hour-glass shaped magnetic field in a HMC with the symmetry axis roughly coinciding with the plane of the hub. This is consistent with the predictions of Kumar et al.~(\cite{kumar20}, \cite{kumar21}), who propose that the accretion of material and the density increase in the hub would compress the initial/local magnetic field. This would increase the magnetic field strength, which in the G31 HMC is of the order of $\sim$10\,mG (Beltr\'an et al.~\cite{beltran19}), stabilize the hub against multiple fragmentation into low-mass objects, and favour fragmentation into fewer high-mass objects with similar masses, as observed in G31 (Beltr\'an et al.~\cite{beltran21}).

\section{Conclusions}

G31 represents a typical example of a hub-filament system and a benchmark for the filaments-to-clusters paradigm of star formation. IRAM\,30m N$_2$H$^+$ observations suggest that star formation in the G31 cloud has started by a cloud-cloud collision, which has led to the formation of the hub-filament system, and ultimately to cluster formation associated with the hot molecular core. In this massive star formation scenario, the rotation of the G31 HMC observed at scales of a few 1000\,au, would have been induced by shear caused by the cloud-cloud collision at scales of a few pc.

\begin{acknowledgements}

This work is based on observations carried out under project number 049$-$21 with the IRAM 30m telescope. IRAM is supported by INSU/CNRS (France), MPG (Germany) and IGN (Spain). We are grateful to the IRAM 30m staff for their help during the different observing runs.  
V.M.R.\ acknowledges support from the Comunidad de Madrid through the Atracci\'on de Talento Investigador Modalidad 1 (Doctores con experiencia) Grant (COOL: Cosmic Origins Of Life; 2019-T1/TIC-15379).

\end{acknowledgements}

\begin{appendix}


\section{Physical parameters of the gas derived from N$_2$H$^+$}

We used the N$_2$H$^+$ spectroscopy from the Cologne Database for Molecular Spectroscopy (CDMS) catalogue\footnote{\url{https://cdms.astro.uni-koeln.de/classic/catalog}} (M\"uller et al.~\cite{muller01}; Endres et al.~\cite{endres16}) entry 029506 (version 4, April 2014) based on the works by Caselli et al.~(\cite{caselli95}) and Cazzoli et al.~(\cite{cazzoli12}) and the dipole moment from Havenith et al.~(\cite{havenith90}). The $J$=1$-$0 transitions splits into fifteen different hyperfine transitions, due to interactions between the molecular electric field gradient and the electric quadrupole moment of the two nitrogen nuclei. Table \ref{tab:spec} shows information of the hyperfine transitions, obtained from the hyperfine entry from CDMS. The partition function takes into account the $^{14}$N hyperfine splitting.

%
%

\begin{table*}
\caption[] {N$_2$H$^+$ spectroscopy used in the analysis.}
\label{tab:spec}
\tabcolsep 8pt 
\begin{tabular}{c c c c c}
\hline
\multicolumn{1}{c}{Frequency (GHz)} & 
\multicolumn{1}{c}{Transition $J$ $F_{\rm 1}$ $F$} &
\multicolumn{1}{c}{log $I$ (nm$^2$ MHz)} &
\multicolumn{1}{c}{log $A_{\rm ul}$ (s${^-1}$)} &
\multicolumn{1}{c}{$E_{\rm up}$ (K)} 
\\
\hline
\multicolumn{5}{c}{$F_{\rm 1}$=1$-$1} \\
\hline
   93.1716157  & 1$_{1,0}-$0$_{1,1}$ & -3.7697   &  -4.44039 & 4.47 \\     
   93.1719106  & 1$_{1,2}-$0$_{1,1}$ & -3.8803    & -4.51336& 4.47  \\   
   93.1719106  & 1$_{1,2}-$0$_{1,2}$ & -3.1439    &-5.25086 & 4.47  \\ 
   93.1720477  & 1$_{1,1}-$0$_{1,0}$  & -3.5859   & -4.73351& 4.47  \\ 
   93.1720477  & 1$_{1,1}-$0$_{1,1}$ &  -4.2155 & 	-4.87021   & 4.47  \\ 
   93.1720477   & 1$_{1,1}-$0$_{1,2}$ & -3.7225  & 	-5.36401 & 4.47  \\ 
\hline 
\multicolumn{5}{c}{$F_{\rm 1}$=2$-$1} \\
\hline
   93.1734734  & 1$_{2,2}-$0$_{1,1}$ & -3.1439    & -4.51335& 4.47        \\ 
   93.1734734  & 1$_{2,2}-$0$_{1,2}$ & -3.8803    &	-5.25085 & 4.47      \\    
   93.1737699  & 1$_{2,3}-$0$_{1,2}$ & -2.9246    &-4.44038	 & 4.47     \\    
   93.1739640  & 1$_{2,1}-$0$_{1,0}$ & -3.7972    &-4.62870	 & 4.47    \\     
   93.1739640  & 1$_{2,1}-$0$_{1,1}$ & -3.4811    & -4.94540	& 4.47     \\    
   93.1739640  & 1$_{2,1}-$0$_{1,2}$ & -4.6978    & -5.84540	& 4.47   \\     
\hline 
\multicolumn{5}{c}{$F_{\rm 1}$=0$-$1} \\
\hline
   93.1762595 & 1$_{0,1}-$0$_{1,0}$  & -4.0417    &-4.67019	 & 4.47   \\     
   93.1762595 & 1$_{0,1}-$0$_{1,1}$  & -3.9256   & -5.07359	 & 4.47    \\     
   93.1762595  & 1$_{0,1}-$0$_{1,2}$  & -3.5224   & -5.18909	& 4.47    \\   
\hline
\end{tabular}
\\
\end{table*}

\begin{table*}
\caption[] {LTE analysis of the molecular emission toward selected regions (see Fig. \ref{fig-n2h+})$^a$.}
\label{table-n2h+}
\tabcolsep 4pt 
\begin{tabular}{c c  ccccc c }
\hline
\multicolumn{1}{c}{Region} 
 & &$N$ & $T_{\rm ex}$  & $FWHM$  & $v_{\rm LSR}$  & $\tau$(1$_{2,3}-$0$_{1,2}$)  \\
 & & ($\times$10$^{12}$ cm$^{-2}$) &  (K) &(km s$^{-1}$) &  (km s$^{-1}$) &    \\ 
\hline
F1 & & 4.1$\pm$1.4 & 3.34$\pm$0.18 & 0.94$\pm$0.07  & 97.83$\pm$0.03 & 0.66$\pm$0.12    \\
F2 & & 3.72$\pm$0.11 & 3.5$^{b}$ & 1.65$\pm$0.09 & 97.25$\pm$0.04 & 0.32$\pm$0.06    \\
P1 & & 24$\pm$3 &  4.23$\pm$0.14  & 2.74$\pm$0.12 & 95.64$\pm$0.06 & 0.9$\pm$0.3    \\
  & & 150$\pm$17  & 3.79$\pm$0.04 & 2.76$\pm$0.13 & 99.09$\pm$0.06 & $>$7    \\
P2 & & 20.9$\pm$1.6  & 6.4$\pm$0.7 & 3.00$\pm$0.09 &  95.55$\pm$0.04 & 0.39$\pm$0.08   \\
F3 && 9$\pm$3 & 3.33$\pm$0.14 & 3.23$\pm$0.13 & 96.06$\pm$0.06 & 0.42$\pm$0.11    \\
F4 & & 6$\pm$3  & 3.4$\pm$0.3 & 3.23$\pm$0.18 & 96.67$\pm$0.07 & 0.3$\pm$0.1    \\

\hline
\end{tabular}
\\
$^{a}$ LTE analysis performed with the AUTOFIT tool of MADCUBA. \\
$^{b}$ Parameter fixed in AUTOFIT.
\end{table*}



\end{appendix}

\end{document}